\documentstyle[epsf,manuscript,aps]{revtex}
\begin{document}
\draft
\preprint{\it Physical Review\/ \rm A \bf 50\rm,
3486--3491 (1994)}
\title{SPIN CORRELATED INTERFEROMETRY FOR\\  
POLARIZED AND UNPOLARIZED PHOTONS\\ ON A BEAM SPLITTER}

\author{Mladen Pavi\v ci\'c}
\address{Institut f\"ur Theoretische Physik, 
TU Berlin, Hardenbergstra\ss e 36, D--10623 Berlin 12, Germany\\
Atominstitut der \"Osterreichischen Universit\"aten, 
Sch\"uttelstra\ss e 115, A--1020 Wien, Austria\\
and Department of Mathematics, University of Zagreb, 
GF, Ka\v ci\'ceva 26, HR--41001 Zagreb, Croatia\cite{mp}}

\date{May 20, 1994}
\maketitle
\widetext
\begin{abstract}
Spin interferometry of the 4th order for independent polarized 
as well as \it unpolarized\/ \rm photons arriving simultaneously at a 
beam splitter and exhibiting spin correlation while leaving it,  
is formulated and discussed in the quantum approach. 
Beam splitter is recognized as a source of genuine 
singlet photon states. 
Also, typical nonclassical beating between photons taking part in 
the interference of the 4th order is given a polarization 
dependent explanation.  
\end{abstract}

\pacs{Ms number AP5065. PACS numbers: 03.65.Bz, 42.50.Wm}

\narrowtext

\section{INTRODUCTION}
\label{sec:intro}
Quite a number of papers were recently engaged in 
the nonclassical 4th order interference of independent 
sources  
\cite{ou88,mand83,paul86,mands87,ozwm90,owzm90,om88,ohm87,ohm88,%
cst89,cst90,yurke92,zeil93,p93,ps93,ps94,wzm91,silv93}.  
It proved to be a powerful tool for checking on possible new 
quantum principles and features as well as on the nonstandard 
interpretations of quantum phenomena. 
E.g., Dirac's principle \it each photon interferes only with itself\/ 
\rm seems to be valid only for the standard interference of the 2nd order 
while for the nonclassical interference of the 4th
order it should read: \it each pair of photons interferes 
only with itself\/\rm\ \cite{paul86,owzm90}. 
As for the nonstandard interpretations, the nonclassical interference of 
the 4th order, in particular with the down--converted beams, was 
recently used for disproving both local and nonlocal hidden--variable 
theories. Ou and Mandel \cite{ou88,om88,ohm88} have elaborated and 
carried out a new type of the Bell--like experiments against local
hidden--variable theories which was then extended by 
Yurke and Stoler \cite{yurke92} to three independent 
sources, by \.{Z}ukowski, Zeillinger, Horne, and Ekert \cite{zeil93} 
to independent correlated pairs in the configuration 
space, and by Pavi\v{c}i\'{c} and Summhammer \cite{p93,ps93} 
to independent correlated pairs in the spin space. 
On the other hand Wang, Zou, and Mandel \cite{wzm91}
carried out an experiment to test de Broglie--Bohm pilot 
(guiding, ``ghost'') waves (without the latter being physically
blocked) according to a set--up proposed by  the 
Selleri--Croca school and obtained a negative result.

The afore mentioned extention in the spin space brought us 
to a new phenomenon --- spin correlated interferometry ---
which asks for an independent elaboration. 
So, in this paper we elaborate the \it spin correlated 
interferometry\/ \rm  of the 4th order for two polarized 
as well as \it unpolarized\/ \rm 
photons arriving simultaneously at a beam splitter and in a 
forthcoming paper \cite{ps94} the spin correlated interferometry 
for the independent pairs of spin (polarization) 
correlated photons. We call the phenomena \it spin correlated 
interferometry \rm because it turns out that two photons which
simulataneously leave a beam splitter, always leave it correlated 
in spin no matter how they were prepared, \it i.e.\rm, no matter whether 
they were previously polarized or not. The interferometry is based on 
an experiment we put forward in Refs.~\cite{p93,ps93} which is a 
realization of the 4th order interference of randomly prepared 
independent photons correlated in polarization and coming from 
independent sources.

Of the two experiments we consider in this paper, the first one  
puts together two polarized photons, makes them interact on a beam 
splitter, and allows us inferring the dependence of the typical 
nonclassical 4th order \it beating \rm on the mutual polarization 
of the incoming photons when no polarization is measured as well as 
inferring modulated polarization (spin) correlations when it is measured. 
The second experiment puts together two unpolarized photons coming out 
from two simultaneous but independent cascade processes of two 
simulataneously excited independent atoms, makes them interact on 
a beam splitter, and then allows us inferring polarization (spin) 
correlations by simultaneous measurement of the polarizations of 
the photons. Thus we recognize beam splitter as a device for detection 
and preparation of spin correlation between incoming and/or outgoing 
photons as well as a source of genuinely unpolarized photons.
This in turn allows us to substitute two additional beam splitters 
for the above cascade sources. 

On the other hand, the present elaboration of the 4th order 
interference on a beam splitter in the spin space attempts to fill a 
gap in the literature. For, while the interference of the 4th order 
in the configuration space has been elaborated in detail in the 
literature \cite{ou88,paul86,mands87,cst89,cst90}, the interference 
lacks a detailed elaboration and apparently a proper understanding in 
the spin space. One of the rare partial elaborations was  
provided by Ou, Hong, and Mandel for a special case of orthogonally 
polarized photons \cite{ohm87b}. They clearly recognized that 
orthogonally polarized photons incoming to a symmetrically positioned 
beam splitter produce a \it singlet--like\/ \rm state at a beam 
splitter \cite{ou88,om88,ohm88,ohm87b} and that parallelly polarized 
photons incoming to a symmetrically positioned beam splitter never 
appear on its opposite sides \cite{hom87} but it does not seem to have 
been recognized that the polarization of incoming photons actually does 
not have any effect on the correlation in polarization of the outgoing 
photons and that it only affects the intensity of 
the outgoing photons. That was also apparently the reason for  
not realizing that a beam splitter produces not a 
\it singlet--like\/ \rm state but a genuine singlet state and 
not only for polarized but for unpolarized incident 
photons as well.   

To be able to follow the main features of the experiments presented 
in Sec.~\ref{sec:exp} we develop the basic formalism in
Sec.~\ref{sec:form} and we shall calculate the setups in the 
plane--wave approach in Sec.~\ref{sec:prob}. 
In Sec.~\ref{sec:concl} we discuss the obtained interference 
patterns. 

\section{EXPERIMENTS}
\label{sec:exp}

The essential part of both experiments --- for polarized as well as 
for unpolarized photons --- is presented in Fig.~\ref{exp} which is,  
in effect, a slightly modified figure from Ref.~\cite{oum89}.  
The only difference for the two cases below are the sources. 

\subsection{Polarized photons}
\label{subsec:pol-ph}

Two incoming independent photons in Fig.~\ref{exp} are emerging 
as signal and idler photons from a nonlinear crystal as in 
the experiment of Ou and Mandel \cite{om88} with the only 
difference that the polarization rotator can be turned in any 
appropriate direction without affecting the output correlation. 
Signal and idler photons of frequences 
$\omega_1$ and $\omega_2$ are produced in the process of parametric 
down conversion of a laser beam (of the frequency 
$\omega_0=\omega_1+\omega_2$) that interacts with the 
nonlinear crystal (\it e.g.\rm, LiIO$_3$). 

The two so obtained independent photons are then directed to a 
beam splitter from opposite sides. Photons coming out from  
the beam splitter pass polarizers P1 and/or P2 and fall on detectors 
D1 and/or D2. In an actual setup a birefringent prisms should 
be used for polarizers (allowing detection of polarization P 
and the perpendicular polarization P$^\perp$) so as to enable 
\it zero detections\/ \rm by appropriate D1$^\perp$,D2$^\perp$ 
detectors (not shown in Fig.~\ref{exp}). 
Pulse pairs arriving within an appropriate time interval 
(typically 5 ns or shorter) are taken as coincidence counts. 
The obtained coincidence counts are ascribed to 
the probability of detecting two photons for possible 
settings of incoming and outgoing (polarizers P1 and P2) 
polarizations. 

Such a setup can however be objected that it fails to adequately 
record photons when they both go to one arm and when their triggering 
of the detectors should be disabled. To match this possibility we can 
use frequency filters (prisms) that would separate the photons emerging 
from the beam splitter according to their frequency and would direct 
them to two birefringent polarizers P$_{\omega_1}$ and P$_{\omega_2}$ 
and through them to four detectors 
D$_{\omega_1}$,D$_{\omega_1}^\perp$,D$_{\omega_2}$,D$_{\omega_2}^\perp$
in each arm. Simultaneous firing (i.e., within the shortest time 
feasible) of at least two detectors in one arm then discards the 
corresponding recording in both arms from the set of \it coincidence
counts\/\rm. In such a way we are able to \it preselect\/ \rm 
a genuine singlet state of the photon pair emerging from 
different sides of the beam splitter (see Sec.~\ref{subsubsec:arms}). 
The setup also enables an experimental verification of the 
behavior of photons when they both emerge from the same side of 
a beam splitter presented in Sec.~\ref{subsubsec:arm}.

\subsection{Unpolarized photons}
\label{subsec:unpol-ph}

Signal and idler down--converted photons emerging from independent 
nonlinear crystals are parallelly polarized ($\pm2^{\circ}$ relative 
to the uv pump laser beam \cite{om88}). So, such sources cannot be 
used to obtain unpolarized incoming photons but we have at least two 
available sources. One is a cascade process, e.g., 
$(J=0)\rightarrow(J=1)\rightarrow(J=0)$. It can be triggered by a 
simultaneous pumping of a split laser beam. Due to the random 
phases of photons emitted from two distinct atoms we shall have no
interference of the 2nd order at all. In previous experiments where 
sources of unpolarized photons were needed, the sources were poorly 
localized because a better localization was not necessary. E.g., in 
Aspect's experiments \cite{asp82} the source atoms where located in a  
$60\times60\>\mu$m region, i.e.~within the laser beam waist 
diameter of the focused pumping laser beam. 
Recently, however, trapping of single atoms (as opposed to 
$3\times 10^{10}$ atoms/cm$^3$ in Aspect's experiment) down to 
$1\times1\>\mu$m has been achieved \cite{eic93}. 

The other possible sources are two other beam splitters. For, 
as will see below, beam splitters emit completely unpolarized 
photons in particular directions. Of course, the beam splitters 
would be much easier to handle than cascade processes but we 
introduced the cascade sources first because of the conceptual clarity 
they give to the proposal of the experiment.    

The rest of the experiment is the same as above for polarized 
photons. 

\section{FORMALISM}
\label{sec:form}

The state of polarized photons immediately after leaving the sources 
is described by the product of two prepared linear--polarization 
states: 
\begin{eqnarray}
|\Psi\rangle=&&\left(\cos\theta_{1'}|1_x\rangle_1\>+
\>\,\sin\theta_{1'}|1_y\rangle_1\right)\nonumber\\
&&\otimes\left(\cos\theta_{2'}|1_x\rangle_2\>+
\>\,\sin\theta_{2'}|1_y\rangle_2\right)\,,\label{eq:2-state}
\end{eqnarray}
where $|1_x\rangle$ and $|1_y\rangle$ denote the mutually orthogonal 
photon states. So, e.g., $|1_x\rangle_1$ means the state of a 
photon leaving the upper source polarized in  
direction $x$. If the beam spliter were removed it would cause 
a \it ``click''\/ \rm at the detector D1 and no \it ``click''\/ \rm 
at the detector D1$^\perp$ provided the birefringent polarizer P1 is 
oriented along $x$. Here D1$^\perp$ means a detector counting 
photons coming out at the \it other exit\/ \rm P$^\perp$
(perpendicular polarization; not shown in the Fig.~\ref{exp}) of the 
birefringent prism P1. Angles $\theta_{1'}$,$\theta_{2'}$ are the angles 
along which incident photons are polarized with respect to a fixed
direction. 

For unpolarized photons the density matrix is proportional 
to the unit matrix and this means that we only need products 
$|1_x\rangle_1\>|1_x\rangle_2$, 
$|1_x\rangle_1\>|1_y\rangle_2$, 
$|1_y\rangle_1\>|1_y\rangle_2$, and 
$|1_y\rangle_1\>|1_y\rangle_2$ to form partial probabilities 
which then sum up to the total correlation probability as shown in 
Sec.~\ref{sec:prob}. 

To describe the interaction of photons with the beam spliter, 
polarizers and detectors we use the quantized electric field operators 
often employed in quantum optical analysis, e.g., 
by Paul \cite{paul86}, Mandel's group \cite{ozwm90,owzm90,ohm87}, 
and Campos et al.~[10].  
Because we use independent sources, resulting random constant phases 
will give no interference of the 2nd order so that we dispense with 
them. As for polarization we introduce it by means of two orthogonal 
scalar field components. Thus the scalar components of the 
stationary electric field operators read: 
\FL
\begin{equation}
\hat E_j({\bf r}_j,t)=
{1\over\sqrt{\cal V}}\sum\limits_{\{\omega_j\}}
l(\omega_j)\hat a(\omega_j)\xi(\omega_j)
e^{i\scriptstyle{\bf k}_j
\cdot\scriptstyle{\bf r}_j-i\omega_jt}\,, 
\end{equation}
where $l(\omega)=i\sqrt{\hbar\omega\over2\varepsilon_\circ}$, 
\bf k \rm is the wave vector ($k=\omega/c$), $j=1,2$ refer to a 
particular photon in question, $\cal V$ is the
quantization volume, $\{\omega\}$ is the frequency set with a
bandwidth $\Delta\omega$, $\hat a(\omega)$ is the annihilation
(lowering) operator at the angular frequency $\omega$, and 
$\xi(\omega)$ is the frequency density of the chosen form for 
the wave packet. In a subsequent paper \cite{ps94} we use the 
Gaussian wave packets and therefore we have 
\begin{equation}
\xi(\omega)=\left[2\pi(\Delta\omega)\right]^{-1/4}
\exp{\left[-\left({\omega-\omega_\circ\over 
2\Delta\omega}\right)^2\right]}\,.
\end{equation}

In this paper we consider only monochromatic waves, i.e., 
$\Delta\omega=0$ and $\xi(\omega)=1$. So, we deal here with
plane waves represented by the following field operators:
\begin{equation}
\hat E_j({\bf r}_j,t)=\hat
a(\omega_j)e^{i\scriptstyle{\bf k}_j\cdot
\scriptstyle{\bf r}_j-i\omega_jt}\,. 
\end{equation}
Of course, we tacitly assume that photons must arrive at the 
beam splitter practically simultaneously, i.e.~with appropriate 
short delays. In the plane wave approach we cannot derive the 
conditions under which events gain a particular visibility 
but that does not affect the reasoning here, since only the 
overall visibility is affected by greater delays. 
In Ref.~\cite{ps94} we carry out the appropriate calculations in detail  
using Gaussian wave packets and we show that the experiment is 
feasible.  

The annihilation operators describe joint actions of polarizers, 
beam splitter, and detectors. The operators act on the states as follows: 
${\hat a}_{1x}|1_{x}\rangle_1=|0_{x}\rangle_1$, \ 
${\hat a}_{1x}^{\dagger}|1_{x}\rangle_1=|2_{x}\rangle_1$, \  
${\hat a}_{1x}|0_{x}\rangle_1=0$, etc.

The action of the beam splitter we describe by the 
input annihilation operators $\hat a_{1in}$ and $\hat a_{2in}$
operators and the following output ones:
\begin{eqnarray}
\hat a_{1out}&&=\ t\hat a_{1in}+i\,r\hat a_{2in}\,,\nonumber\\
\hat a_{2out}&&=i\,r\hat a_{1in}+\ t\hat a_{2in}\,,
\end{eqnarray}
where $t=|\sqrt T|$ and $r=|\sqrt R|$, where $T$ and $R$ denote 
transmittance and reflectance, respectively. 

To take the linear polarization along orthogonal directions into
account we shall consider two sets of operators, i.e., their matrices 
\begin{equation}
\hat{\bf a}_{x\,out}={\bf c}\,\hat{\bf a}_{x\,in} \ \ \ {\rm and} 
 \ \ \ \hat{\bf a}_{y\,out}={\bf c}\,\hat{\bf a}_{y\,in}\,.
\end{equation}

So, the action of the polarizers P1,P2 and detectors D1,D2 can 
be expressed as: 
\begin{equation}
{\hat a}_i={\hat a}_{ix\,out}\cos\theta_i+
{\hat a}_{iy\,out}\sin\theta_i\,,\label{eq:D2} 
\end{equation}
where $i=1,2$. 

Projections corresponding to the other choices of polarizers and 
detectors we obtain by using appropriate transformations instead of 
the ones given by Eqs.~(\ref{eq:e1}) and (\ref{eq:e2}). 
E.g., we obtain the action of the polarizer P2$^\perp$ (orthogonal 
to P2; in the experiment P2 and P2$^\perp$ make a birefringent prism) 
and the corresponding detector D2$^\perp$ if we substitute  
\begin{equation}
{\hat a}_2=-{\hat a}_{2x\,out}\sin\theta_2+
{\hat a}_{2y\,out}\cos\theta_2\label{eq:D2'-perp} 
\end{equation}
for Eq.~(\ref{eq:D2}).

Hence the appropriate outgoing electric field operators read
\FL
\begin{eqnarray}
\hat E_1&&=\left(\hat a_{1x}t_x\cos\theta_1+\hat
a_{1y}t_y\sin\theta_1\right)
e^{i\scriptstyle{\bf k}_1\cdot
\scriptstyle{\bf r}_1-
i\omega_1(t-\tau_1)}\nonumber\\
+&&i\left(\hat a_{2x}r_x\cos\theta_1+\hat
a_{2y}r_y\sin\theta_1\right)
e^{i\tilde{\scriptstyle{\bf k}}_2\cdot
\scriptstyle{\bf r}_1-
i\omega_2(t-\tau_2)}\,,\label{eq:e1}
\end{eqnarray}
\FL
\begin{eqnarray}
\hat E_2&&=\left(\hat a_{2x}t_x\cos\theta_2+\hat
a_{2y}t_y\sin\theta_2\right)e^{i\scriptstyle{\bf k}_2
\cdot\scriptstyle{\bf r}_2-i\omega_2(t-\tau_2)}\nonumber\\
+&&i\left(\hat a_{1x}r_x\cos\theta_2+\hat
a_{1y}r_y\sin\theta_2\right)e^{i\tilde{\scriptstyle{\bf k}}_1\cdot
\scriptstyle{\bf r}_2-i\omega_1(t-\tau_1)}\,,\label{eq:e2}
\end{eqnarray}
where $\tau_{j}$ is time delay after which the photon reaches detector
D, $\omega_{j}$ is the frequency of photon $j$, and $c$ is the velocity 
of light. The detectors and the crystal are assumed to be positioned
symmetrically with regard to the beam splitter so that two time 
delays suffice.

\section{Detection probabilities}
\label{sec:prob}

\subsection{Polarized photons}
\label{subsec:pol-prob}

\subsubsection{Each photon in one arm}
\label{subsubsec:arms}

The joint interaction of both photons with the beam splitter, 
polarizers P1,P2, and detectors D1,D2 is given by the following
projection of our wave function onto the Fock vacuum space 
\widetext
\FL
\begin{eqnarray}
\hat E_1\hat E_2|\Psi\rangle=&&\Bigl[\bigl(t_x^2\varepsilon_{12}-
r_x^2\tilde\varepsilon_{12}\bigr)
\cos\theta_{1'}\cos\theta_{2'}\cos\theta_1\cos\theta_2\nonumber\\
&&+\bigl(t_x t_y \varepsilon_{12}\sin\theta_1\cos\theta_2-
r_x r_y \tilde\varepsilon_{12}\cos\theta_1\sin\theta_2\bigr)
\sin\theta_{1'}\cos\theta_{2'}\nonumber\\
&&+\bigl(t_x t_y\varepsilon_{12}\cos\theta_1\sin\theta_2-
r_x r_y \tilde\varepsilon_{12}\sin\theta_1\cos\theta_2\bigr)
\cos\theta_{1'}\sin\theta_{2'}\nonumber\\
&&+\bigl(t_y^2\varepsilon_{12}-r_y^2\tilde\varepsilon_{12}\bigr)
\sin\theta_{1'}\sin\theta_{2'}\sin\theta_1\sin\theta_2\Bigr]\varepsilon
|0\rangle\,,\label{eq:proj}
\end{eqnarray}
where $\varepsilon=\exp\Bigl\{-i\left[\omega_1\left(t-\tau_1\right)+
\omega_2\left(t-\tau_2\right)\right]\Bigr\}$,
$\varepsilon_{12}=\exp\bigl[i\left(
{\bf k}_1\cdot{\bf r}_1+
{\bf k}_2\cdot{\bf r}_2\right)\bigr]$, 
and $\tilde \varepsilon_{12}=\exp\left[i\left(
\tilde{{\bf k}}_1\cdot{\bf r}_2+
\tilde{{\bf k}}_2\cdot{\bf r}_1
\right)\right]$.  

The corresponding probability of detecting the photons by
detectors D1,D2 is thus
\begin{equation}
P(\theta_{1'},\theta_{2'},\theta_1,\theta_2)
=\langle\hat E_2^\dagger\hat E_1^\dagger
\hat E_1\hat E_2\rangle
=A^2+B^2-2AB\cos\phi\,,\label{eq:prob} 
\end{equation}
where 
\begin{eqnarray}
A=t_x^2&&\cos\theta_{1'}\cos\theta_{2'}\cos\theta_1\cos\theta_2+
t_y^2\sin\theta_{1'}\sin\theta_{2'}\sin\theta_1\sin\theta_2\nonumber\\
&&+t_x t_y\left(\cos\theta_{1'}\sin\theta_{2'}\cos\theta_1\sin\theta_2+
\sin\theta_{1'}\cos\theta_{2'}\sin\theta_1\cos\theta_2\right)\,,
\end{eqnarray}
\begin{eqnarray}
B=r_x^2&&\cos\theta_{1'}\cos\theta_{2'}\cos\theta_1\cos\theta_2+
r_y^2\sin\theta_{1'}\sin\theta_{2'}\sin\theta_1\sin\theta_2\nonumber\\
&&+r_x r_y\left(\cos\theta_{1'}\sin\theta_{2'}\sin\theta_1\cos\theta_2+
\sin\theta_{1'}\cos\theta_{2'}\cos\theta_1\sin\theta_2\right)\,,
\end{eqnarray}
\begin{equation}
\phi=(\tilde{{\bf k}}_2-
{\bf k}_1)\cdot{\bf r}_1+
(\tilde{{\bf k}}_1-
{\bf k}_2)\cdot{\bf r}_2 
=2\pi(z_2-z_1)/L\,,
\end{equation}
\narrowtext
\noindent where $L$ is the spacing of the intereference fringes 
\cite{ou88}. $\phi$ can be changed by moving the detectors 
transversely to the incident beams.

To make the formula more transparent, without loss of generality, 
in the following we shall consider 50:50 beam splitter: 
$t_x=t_y=r_x=r_y=2^{-1/2}$ and three characteristic locations of 
the detectors so as to have $\cos\phi=-1,0,1$ 

Let us first consider the case $\phi=0$ for which the above 
probability reads
\begin{eqnarray}
P(\theta_{1'},\theta_{2'},\theta_1,\theta_2)=&&(A-B)^2\nonumber\\
=&&{1\over4}\sin^2(\theta_{1'}-\theta_{2'})\sin^2(\theta_1-\theta_2)\,.
\label{eq:coinc} 
\end{eqnarray}
We see that the probability unexpectedly factorizes left--right 
and not up--down as one would be tempted to conjecture from the 
initial up--down independence expressed by the product of the 
``upper'' and ``lower'' function in Eq.~(\ref{eq:2-state}). 

On the other hand, the incoming polarizations influence the 
coincidence counting even when we remove the polarizers P1 and P2. 
Then, provided the right photons arrive at the
beam splitter within a sufficiently short time 
and are separately detected by D1 and D2, we obtain  
\begin{equation}
P(\theta_{1'},\theta_{2'},\infty,\infty)=
{1\over2}\sin^2(\theta_{1'}-\theta_{2'})\,.\label{eq:bingo} 
\end{equation}

This equation clarifies the minimum of the coincidence rates 
obtained for the $z_1=z_2$ positions of detectors in 
Refs.~\cite{oomm88,ohmm87,rt89,omfig89}. 
We just have to recall again that signal and idler down--converted 
photons emerging from independent nonlinear crystals used in these 
experiments are parallelly polarized \cite{om88}. Conversely, by 
inserting $\theta_{2'}=\theta_{1'}+\pi/2$ in Eq.~(\ref{eq:coinc}) 
we obtain exactly what --- for $\phi=0$ --- Ou, Hong, and Mandel 
obtained in Ref.~\cite{ohm88} and what Ou and Mandel should have 
obtained also in Refs.~\cite{ou88} and \cite{om88}. \cite{no-ou}  

For $\phi=\pi$ our probability reads 
\FL
\begin{eqnarray}
&&P(\theta_{1'},\theta_{2'},\theta_1,\theta_2)=(A+B)^2\\ 
&&={1\over4}[
\cos(\theta_{1'}\!-\!\theta_2)\cos(\theta_{2'}\!-\!\theta_1)\!+\! 
\cos(\theta_{1'}\!-\!\theta_1)\cos(\theta_{2'}\!-\!\theta_2)]^2\!,
\nonumber 
\end{eqnarray}
while for $\phi=\pi/2$ it becomes
\FL
\begin{eqnarray}
P(\theta_{1'},\theta_{2'},\theta_1,\theta_2)=&& 
{1\over4}\bigl[
\cos^2(\theta_{1'}-\theta_2)\cos^2(\theta_{2'}-\theta_1)\nonumber\\ 
&&+\cos^2(\theta_{1'}-\theta_1)\cos^2(\theta_{2'}-\theta_2)\bigr]\,.
\end{eqnarray}
The probability shows that, for $\phi=\pi$, 
by removing the polarizers we lose the spin correlation 
completely and the coincidence counting remains unchanged no matter 
how we turn the polarization planes of the incoming photons. 
This is just opposite to $\phi=0$ above where because of 
Eq.~(\ref{eq:bingo}) we could not have a coincidence for 
parallel incident polarizations. The latter means that we obtain 
the typical non--classical 100\%\ (ideally) coincidence rate 
\cite{oomm88,ohmm87,rt89} as opposed to the classical treatment 
(maximum 50\%), i.e., that both photons go into  
only one of the arms. Let us therefore have a closer 
look at the case of two photons in a particular arm.

\subsubsection{Both photons in one arm}
\label{subsubsec:arm}

In order to treat both photons going into one arm  
properly (i.e., so as to make all the probabilities add up to one) 
we have to switch to the experimental setup described in the 
last paragraph of Sec.~\ref{subsec:pol-ph} and employ
four detectors in each arm: D2$_{\omega_1}$--D2$_{\omega_2}^\perp$ 
and D1$_{\omega_1}$--D1$_{\omega_2}^\perp$ in the upper and lower 
arm, respectively. Let us do that for the upper arm.
Instead of $\hat E_1$ from Eq.~(\ref{eq:e1}) we must use 
\FL
\begin{eqnarray}
\hat E_2'&&=\left(\hat a_{2x}t_x\cos\theta_1+\hat
a_{2y}t_y\sin\theta_1\right)
e^{i\scriptstyle{\bf k}_2'\cdot\scriptstyle{\bf r}_2'-
i\omega_2(t-\tau_2)}\nonumber\\
+&&i\left(\hat a_{1x}r_x\cos\theta_1+\hat
a_{1y}r_y\sin\theta_1\right)
e^{i\tilde{\scriptstyle{\bf k}}_1'\cdot
\scriptstyle{\bf r}_2'-
i\omega_1(t-\tau_1)}\label{eq:e1-two}
\end{eqnarray}
so as to obtain the following analogue of Eq.~(\ref{eq:proj})
\widetext
\FL
\begin{eqnarray}
\hat E_2'\hat E_2|\Psi\rangle=&&\Bigl[t_xr_x\bigl(\eta_{2'}+\eta_2\bigr)
\cos\theta_{1'}\cos\theta_{2'}\cos\theta_1\cos\theta_2\nonumber\\
&&+t_xr_y\bigl(\eta_{2'}\cos\theta_1\sin\theta_2+
\eta_2\sin\theta_1\cos\theta_2\bigr)
\sin\theta_{1'}\cos\theta_{2'}\nonumber\\
&&+t_yr_x\bigl(\eta_{2'}\sin\theta_1\cos\theta_2+
\eta_2\cos\theta_1\sin\theta_2\bigr)
\cos\theta_{1'}\sin\theta_{2'}\nonumber\\
&&+t_yr_y\bigl(\eta_{2'}+\eta_2\bigr)
\sin\theta_{1'}\sin\theta_{2'}\sin\theta_1\sin\theta_2\Bigr]\varepsilon
|0\rangle\,,\label{eq:projx}
\end{eqnarray}
where $\varepsilon=\exp\Bigl[-i\left[\omega_1\left(t-\tau_1\right)+
\omega_2\left(t-\tau_2\right)\right]\Bigr]$,
$\eta_{2'}=\exp\bigl[i\left(
{\bf k}_1\cdot{\bf r}_2+
{\bf k}_2'\cdot{\bf r}_2'\right)\bigr]$, 
and $\eta_2=\exp\left[i\left(\tilde{{\bf k}}_2
\cdot{\bf r}_2+\tilde{{\bf k}}_1'\cdot
{\bf r}_2'\right)\right]$.  

The corresponding probability of detecting the photons by
detectors D2$_{\omega_1}$,D2$_{\omega_2}$ is thus
\begin{eqnarray}
P(\theta_{1'},\theta_{2'},\theta_1\times\theta_2)
={1\over2}\langle\hat E_2^{\dagger}\hat E_2'^{\dagger}
\hat E_2'\hat E_2^{\phantom\dagger}\rangle
={1\over2}(C^2+D^2+2CD\cos\psi)\,,\label{eq:prob-2} 
\end{eqnarray}
where 1/2 matches the possibility of both photons taking 
the other arm and 
\begin{eqnarray}
C=t_xr_x&&\cos\theta_{1'}\cos\theta_{2'}\cos\theta_1\cos\theta_2+
t_yr_y\sin\theta_{1'}\sin\theta_{2'}\sin\theta_1\sin\theta_2\nonumber\\
&&+t_xr_y\sin\theta_{1'}\cos\theta_{2'}\sin\theta_1\cos\theta_2+
t_yr_x\cos\theta_{1'}\sin\theta_{2'}\cos\theta_1\sin\theta_2\,,
\end{eqnarray}
\begin{eqnarray}
D=t_xr_x&&\cos\theta_{1'}\cos\theta_{2'}\cos\theta_1\cos\theta_2+
t_yr_y\sin\theta_{1'}\sin\theta_{2'}\sin\theta_1\sin\theta_2\nonumber\\
&&+t_x r_y\sin\theta_{1'}\cos\theta_{2'}\cos\theta_1\sin\theta_2+
t_yr_x\cos\theta_{1'}\sin\theta_{2'}\sin\theta_1\cos\theta_2\,,
\end{eqnarray}
\begin{equation}
\psi=(\tilde{{\bf k}}_1-{\bf k}_2)\cdot{\bf r}_2+
({\bf k}_2'-\tilde{{\bf k}}_1')\cdot{\bf r}_2' 
=2\pi(Z_2-Z_2')/L\,,\label{eq:2-arms}
\end{equation}
\narrowtext
\noindent where primes refer to the \it other\/ \rm photon of a 
different frequency and where the geometry of the detectors is 
of course not any more the one shown in Fig.~\ref{exp} but is, e.g., 
following the Fig.~1 of Ref.~\cite{gryu88}. An analogous probability 
we obtain for the lower arm.

For 50:50 beam splitter and $\psi=0$ the probability reads 
\FL
\begin{eqnarray}
&&P(\theta_{1'},\theta_{2'},\theta_1\times\theta_2)\\ 
&&={1\over8}[
\cos(\theta_{1'}\!-\!\theta_2)\cos(\theta_{2'}\!-\!\theta_1)\!+\! 
\cos(\theta_{1'}\!-\!\theta_1)\cos(\theta_{2'}\!-\!\theta_2)]^2\!.
\nonumber 
\end{eqnarray}

To obtain the corresponding probability with the polarizers
removed we have to add up probabilities for all
four possible outcomes from the birefringent P2 which we obtain by 
using Eqs.~(\ref{eq:D2}),~(\ref{eq:D2'-perp}), and two other ones 
what for both arms amounts to 
\begin{equation}
P(\theta_{1'},\theta_{2'},\infty\times\infty)=
{1\over2}\bigl[1+\cos^2(\theta_{1'}-\theta_{2'})\bigr]. 
\end{equation}
We see that this equation and Eq.~(\ref{eq:bingo}) add up to one.

Another possible way of detecting both photons in one arm,
although far less reliable, is by means of \it non--coincidental\/
\rm recording of only one of detectors D1,D2 assuming the recording
be triggered by two simultaneously arriving photons.
In this case we keep to the setup described in the second
paragraph of Sec.~\ref{subsec:pol-ph} and employ
no additional detectors.
Then the  probability of detecting both photons in the arm of, e.g.,
D2 we obtain similarly to Eq.~(\ref{eq:prob})
\FL
\begin{eqnarray}
P&&(\theta_{1'},\theta_{2'},2\times\theta_2)
={1\over2}\langle\hat E_2^{\dagger2}\hat E_{2}^2\rangle\nonumber\\
&&={1\over8}\bigl[\cos^2(\theta_{1'}-\theta_2)
\cos^2(\theta_{2'}-\theta_2)\bigr](1+\cos\psi)\,,\label{eq:twoarms}
\end{eqnarray}
where $\psi$ is automatically zero because of the coincidental
spatial recording of both photons. Of course, we cannot add up this 
probability for the removed polarizers and the probability 
(\ref{eq:bingo}) to 1 because the the corresponding counts are from 
two different spaces of events.

\subsection{Unpolarized photons}
\label{subsec:unpol-prob}

To obtain the general probability for unpolarized light,  
$\hat E_1,\hat E_2$  given by Eqs.~(\ref{eq:e1}) and 
(\ref{eq:e2}) should be applied to $|1_x\rangle_1|1_x\rangle_2$, 
$|1_x\rangle_1|1_y\rangle_2$, $|1_y\rangle_1|1_x\rangle_2$, 
and $|1_y\rangle_1|1_y\rangle_2$ so as to give four probabilities 
which then sum up to the following correlation probability:

\FL
\begin{eqnarray}
&&P(\infty,\infty,\theta_1,\theta_2)\nonumber\\
&&={1\over4}(t_x^2\cos^2\theta_1+t_y^2\sin^2\theta_1)
(t_x^2\cos^2\theta_2+t_y^2\sin^2\theta_2)\nonumber\\
&&+{1\over4}(r_x^2\cos^2\theta_1+r_y^2\sin^2\theta_1)
(r_x^2\cos^2\theta_2+r_y^2\sin^2\theta_2)\nonumber\\
&&-{1\over2}(t_x^2r_x^2\cos\theta_1\cos\theta_2+
t_y^2r_y^2\sin\theta_1\sin\theta_2)^2\cos\phi\>.
\label{eq:bangg}
\end{eqnarray}

For 50:50 beam splitter this probability reads
\begin{equation}
P(\infty,\infty,\theta_1,\theta_2)=  
{1\over8}\bigl[1-\cos\phi\cos^2 (\theta_2-\theta_1)\bigr]\,.
\label{eq:bang}
\end{equation}

Comparing this result with the classical formula obtained by 
Paul \cite{paul} for two amplitude--stabilized beams of equal
intensity which, apart from a normalization factor, reads 
\begin{equation}
P_{\mbox{\rm\tiny cl}} (\theta_1,\theta_2) =
3+2 (1-\cos\phi)\cos^2(\theta_2 -\theta_1)
\,, 
\end{equation}
we see that the quantum mechanical \it visibility\/ \rm reaches 
its maximum for $\phi=0$ while the corresponding classical one 
cannot be equal to zero at all. 

In the end, for unpolarized photons and for $\phi=0$ we obtain:
\begin{equation}
P(\infty,\infty,\theta_1,\theta_2)=
{1\over8}\sin^2(\theta_2-\theta_1)\,.\label{eq:emerge} 
\end{equation}
So, photons that arrive at the beam splitter unpolarized emerge 
from it perpendicularly polarized whenever they appear at the 
opposite sides of the beam splitter. The overall probability of 
their appearance on one side of the beam splitter is 
\begin{equation}
P(\infty,\infty,\theta_1\times\theta_2)=
{1\over8}\bigl[1+\cos^2(\theta_1-\theta_2)\bigr]. 
\end{equation}

\section{CONCLUSION}
\label{sec:concl}

We have therefore shown that the 4th order interference interaction 
between beam splitter and two incoming photons imposes 
polarization correlation on the emerging photons no matter whether 
they arrive at the beam splitter polarized or unpolarized. In particular 
we have shown [Eqs.~(\ref{eq:bang}) and (\ref{eq:emerge})] that for an 
appropriate position of the beam splitter incoming unpolarized photons 
always emerge perpendicularly polarized in particular directions.
More specifically, they appear prepared in a genuine singlet state 
and enable conceiving a novel experiment in which we can preselect 
spin correlated photons from completely unpolarized independent 
photons which nowhere interacted without in any way affecting 
them \cite{p94}.

When polarized photons arrive at a beam splitter and the 4th order 
interference takes place, one can use the modulation of the 
polarizations in order to determine the coincidence counting even 
when no outgoing polarization is being measured. 
In particular we have shown [Eq.~(\ref{eq:bingo})] that in 
predetermined directions incoming parallelly
polarized photons never emerge on two different sides of the 
beam splitter. It is also interesting that the probability of 
detecting both photons together on one side of the beam splitter 
(by one detector) is structurally different from the the 
probability of finding them on both sides. 
The former depends on the direction of leaving the beam splitter 
and allows a transmission of the left--right information 
of the Bell type [Eq.~(\ref{eq:twoarms})].

\acknowledgments
I am grateful to Dr.~Summhammer for particularly fruitful and 
rich collaboration and for his hospitality at the Atominstitut 
der \"Osterreichischen Universit\"aten, 
Vienna, to Prof.~K.--E.~Hellwig for valuable discussions and the 
hospitality at the Institut f\"ur Theoretische Physik, TU Berlin 
and to Prof.~H.~Paul, Humboldt--Universit\"at zu Berlin for valuable 
discussions and correspondence and a pronounced active interest for 
the idea developed in the papers \cite{p93,ps93,ps94}. I also 
acknowledge supports of the Alexander von Humboldt Foundation, the 
Technical University of Vienna, and the Ministry of Science of Croatia.

\begin{figure}
\caption{Outline of the experiment.}
\label{exp} 
\end{figure}

%\vfill\eject 

%\vfill 

\vbox to 2cm{\vfill}

\epsffile{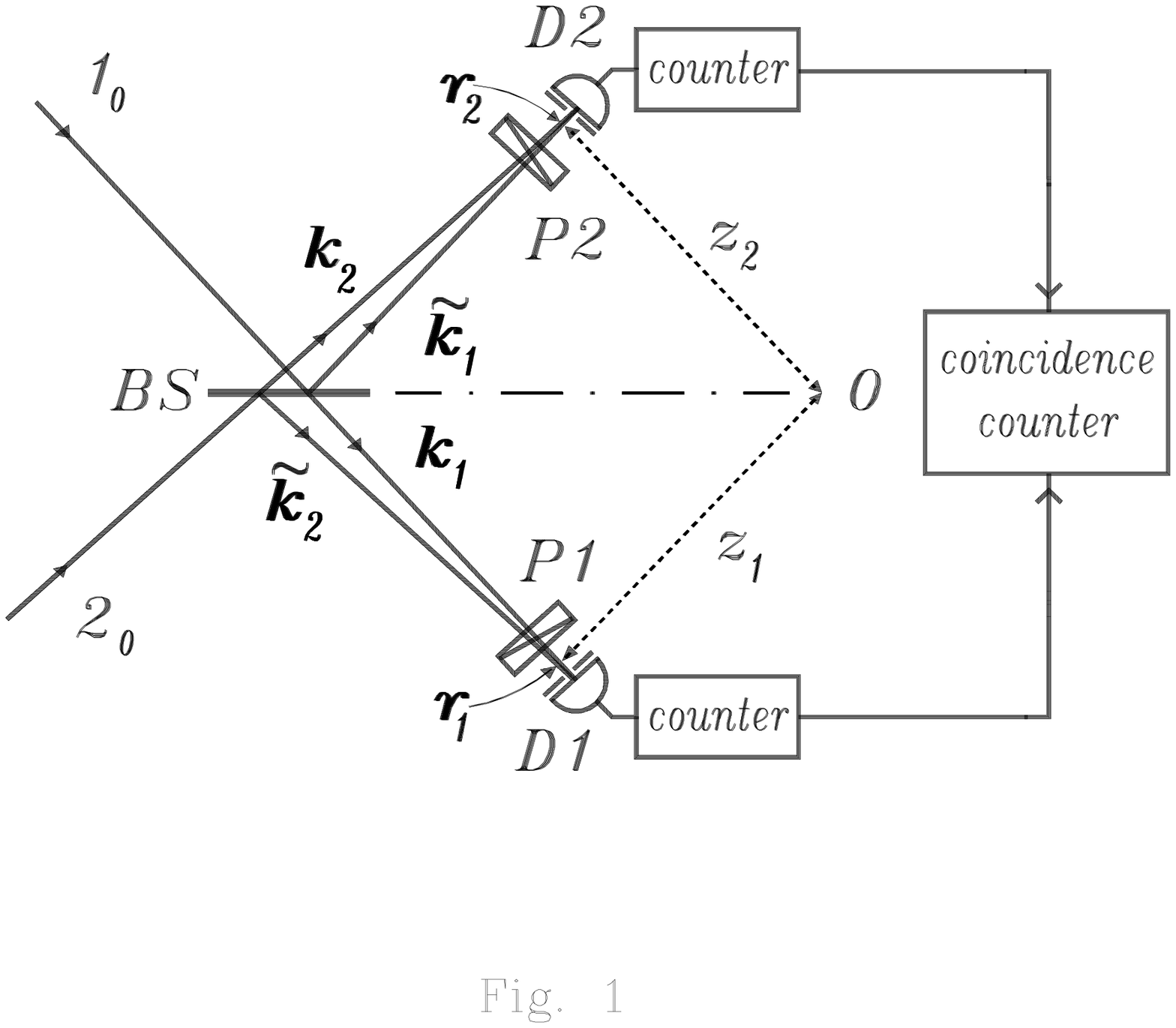}


\begin{references}

\bibitem[*]{mp}  Permanent address. Internet: mpavicic@faust.irb.hr; 
                    Web: http://m3k.grad.hr/pavicic. 

\bibitem{ou88}   Z. Y. Ou, 
                    Phys.\ Rev.\ A {\bf 37}, 1607 (1988).
\bibitem{mand83} L. Mandel, 
                    Phys.\ Rev.\ A {\bf 28}, 929 (1983). 
\bibitem{paul86} H. Paul, 
                    Rev.\ Mod.\ Phys.\ {\bf 57}, 209 (1986).
\bibitem{mands87}R. Gosh and L. Mandel, 
                    Phys.\ Rev.\ Lett.\ {\bf 59}, 1903 (1987); 
                 Y. H. Shih and C. O. Alley, 
                    Phys.\ Rev.\ Lett.\ {\bf 61}, 2921 (1988);  
                 J. D. Franson, 
                    Phys.\ Rev.\ Lett.\ {\bf 62}, 2205 (1989); 
                 J. G. Rarity et al. 
                    Phys.\ Rev.\ Lett.\ {\bf 65}, 1348 (1990); 
                 Z. Y. Ou et al.    
                    Phys.\ Rev.\ Lett.\ {\bf 65}, 321 (1990); 
                 X. Y. Zou, L. J. Wang, and L. Mandel, 
                    Phys.\ Rev.\ Lett.\ {\bf 67}, 318 (1991); 
                 X. Y. Zou, T. P. Grayson, and L. Mandel, 
                    Phys.\ Rev.\ Lett.\ {\bf 69}, 3041 (1992); 
                 X. Y. Zou et al. 
                    Phys.\ Rev.\ A {\bf 47}, 2293 (1993).  
\bibitem{ozwm90} Z. Y. Ou,  X. Y. Zou, L. J. Wang, and L. Mandel, 
                    Phys.\ Rev.\ A {\bf 42}, 2957 (1990).
\bibitem{owzm90} Z. Y. Ou, L. J. Wang, X. Y. Zou, and L. Mandel, 
                    Phys.\ Rev.\ A {\bf 41}, 566 (1990).
\bibitem{om88}   Z. Y. Ou and L. Mandel, 
                    Phys.\ Rev.\ Lett.\ {\bf 61}, 50 (1988); 
\bibitem{ohm87}  X. Y. Ou, C. K. Hong, and L. Mandel, 
                    Phys.\ Lett.\ A {\bf 122}, 11 (1987).
\bibitem{ohm88}  X. Y. Ou, C. K. Hong, and L. Mandel, 
                    Opt.\ Commun.\ {\bf 67}, 159 (1988).
\bibitem{cst89}  R. A. Campos, B. E. A. Saleh, and M. A. Teich,   
                    Phys.\ Rev.\ A {\bf 40}, 1371 (1989). 
\bibitem{cst90}  R. A. Campos, B. E. A. Saleh, and M. A. Teich,   
                    Phys.\ Rev.\ A {\bf 42}, 4127 (1990). 
\bibitem{yurke92}B. Yurke and D. Stoler, 
                    Phys.\ Rev.\ Lett.\ {\bf 68}, 1251 (1992);
                 B. Yurke and D. Stoler, 
                    Phys.\ Rev.\ A {\bf 46}, 2229 (1992). 
\bibitem{zeil93} M.\.Zukowski, A. Zeilinger, M. A. Horne, and A. K. Ekert, 
                    Phys.\ Rev.\ Lett.\ {\bf 71}, 4287 (1993). 
\bibitem{p93}    M. Pavi\v ci\'c, 
                    \it Interference of four correlated beams from 
                    nonlocal sources\rm, (presented at the Inst. f. 
                    Theor. Phys., Techn. Univ. Berlin, on July 16,
                    1993).
\bibitem{ps93}   M. Pavi\v ci\'c and J. Summhammer, 
                    Phys.\ Rev.\ Lett.\ [submitted] (1993). 
\bibitem{ps94}   M. Pavi\v ci\'c and J. Summhammer, 
                    \it Preselection of spin correlated pairs of photons 
                    from independent sources\/\rm, (unpublished, 1994).  
\bibitem{wzm91}  L. J. Wang, X. Y. Zou, and L. Mandel, 
                    Phys.\ Rev.\ Lett.\ {\bf 66}, 1111 (1991); 
                 See also X. Y. Zou et al., 
                    Phys.\ Rev.\ Lett.\ {\bf 68}, 3667 (1992). 
\bibitem{silv93} M. P. Silverman, 
                    Am.\ J.\ Phys.\ {\bf 61}, 514 (1993).
\bibitem{ohm87b} X. Y. Ou, C. K. Hong, and L. Mandel, 
                    Opt.\ Commun.\ {\bf 63}, 118 (1987).
\bibitem{hom87}  C. K. Hong,  Z. Y. Ou and L. Mandel, 
                    Phys.\ Rev.\ Lett.\ {\bf 59}, 2044 (1987). 
\bibitem{oum89}  C. K. Hong,  Z. Y. Ou and L. Mandel, 
                    Phys.\ Rev.\ Lett.\ {\bf 62}, 1903 (1989); 
\bibitem{asp82}  A. Aspect, J. Dalibard, and G. Roger,  
                    Phys.\ Rev.\ Lett.\ {\bf 49}, 1804 (1982).
\bibitem{eic93}  U. Eichmann et al., 
                    Phys.\ Rev.\ Lett.\ {\bf 70}, 2359 (1993).
\bibitem{oomm88} Z. Y. Ou and L. Mandel, 
                    Phys.\ Rev.\ Lett.\ {\bf 61}, 54 (1988). 
\bibitem{ohmm87} C. K. Hong, Z. Y. Ou and L. Mandel, 
                    Phys.\ Rev.\ Lett.\ {\bf 59}, 2044 (1987).  
\bibitem{rt89}   J. G. Rarity and P. R. Tapster, 
                    Opt.\ Soc.\ Am.\ B {\bf 6}, 1221 (1989). 
\bibitem{omfig89}Z. Y. Ou and L. Mandel, 
                    Phys.\ Rev.\ Lett.\ {\bf 62}, 2941 (1989).  
\bibitem{no-ou}  Eqs.~(56), (57), etc.~of Ref.~\protect\cite{ou88} and 
                    Eqs.~(2), (5a), etc.~of Ref.~\protect\cite{om88}
                    should be corrected so as to change signs in order 
                    to assure the preservation of boson commutation 
                    relations for the annihilation operators at the
                    beam splitter. I would like to mention here that 
                    Prof.~H.~Paul noticed this mistake as well 
                    (in a private communication). 
\bibitem{gryu88} P. Grangier, M. J. Potasek, and B. Yurke, 
                    Phys.\ Rev.\ A {\bf 38}, 3132 (1988). 
\bibitem{paul}   H. Paul, 
                    Polarization correlation in unpolarized 
                    light on beam splitter, 
                    (unpublished, 1993). 
                    Actually, Prof.~Paul's paper triggered the way in 
                    which I presented the matter in the present paper 
                    and he himself was brought to consider the 
                    classical approach through the papers 
                    \protect\cite{p93} and \protect\cite{ps93}. 
\bibitem{p94}    M. Pavi\v ci\'c, 
                    \it Preselection of spin correlated photons\/\rm, 
                                          (unpublished, 1994).

\end{references}
\end{document}